\documentstyle[12pt]{article}
\input amssym.def
\input amssym.tex
\parskip=\medskipamount
\newtheorem{definition}{Definition}[section]

\newtheorem{theorem}[definition]{Theorem}
\newtheorem{proposition}[definition]{Proposition}
\newtheorem{corollary}[definition]{Corollary}
\newtheorem{remark}[definition]{Remark}
\newtheorem{example}[definition]{Example}
\newtheorem{examples}[definition]{Examples}

\font\fr=eufm10  scaled \magstep 1   %Caracteres g"ticos.
\font\ddpp=msbm10  scaled \magstep 1  %Caracteres "doble palo".

\def\QED{\hskip0.1em\hfill\null\ \null\nobreak\hfill
\kern3pt\lower1.8pt\vbox{\hrule\hbox   {\vrule\kern1pt\vbox{\kern1.7pt
\hbox{$\scriptstyle   QED$}\kern0.2pt}\kern1pt\vrule}\hrule}}
\def\R{\hbox{\ddpp R}}               %Numeros reales
\def\lcf{\lbrack\! \lbrack}
\def\rcf{\rbrack\! \rbrack}

\begin{document}
\baselineskip=.55cm
\title{Leibniz algebroid associated with a Nambu-Poisson structure} 
\author{R. IBA\~NEZ $^{1}$, M. de LEON $^2$, J.C. MARRERO $^{3}$ and E.
PADRON $^{3} $ 
\\[10pt]
{\small\it $^1$Departamento de Matem\'aticas, Facultad de Ciencias,}\\[-8pt]
{\small \it Universidad del Pais Vasco,}\\[-8pt]
{\small \it Apartado 644, 48080 Bilbao, Spain,}\\[-8pt]
{\small \it  mtpibtor@lg.ehu.es}\\[8pt]
{\small \it $^2$Instituto de Matem\'aticas y F{\'\i}sica Fundamental,}\\[-8pt]
{\small\it Consejo Superior de In\-ves\-ti\-ga\-cio\-nes
Ci\-en\-t{\'\i}\-fi\-cas,} \\[-8pt]
{\small\it Serrano 123, 28006 Madrid, SPAIN,}\\[-8pt]
{\small\it E-mail: mdeleon@fresno.csic.es} \\[8pt]
{\small\it $^3$Departamento de Matem\'atica Fundamental,
Facultad de Matem\'aticas,}\\[-8pt]
{\small\it Universidad de la Laguna,
La Laguna,} \\[-8pt]
{\small\it Tenerife, Canary Islands, SPAIN,}\\[-8pt]
{\small\it E-mail: jcmarrer@ull.es, mepadron@ull.es }
}
%\date{\today}

\maketitle

\begin{abstract}
The notion of Leibniz algebroid is introduced, and it is shown that
each Nambu-Poisson manifold has associated a canonical Leibniz
algebroid. This fact permits to define the modular class of a
Nambu-Poisson manifold as an appropiate cohomology class, extending the
well-known modular class of Poisson manifolds.
\end{abstract}

\begin{quote}
{\it Mathematics Subject Classification} (1991): 53C15, 58F05, 81S10.

{\it PACS numbers}: 02.40.Ma, 03.20.+i, 0.3.65.-w

{\it Key words and phrases}: 
Nambu-Poisson brackets, Nambu-Poisson manifolds, Leibniz algebras,
Leibniz cohomology, Leibniz algebroid, modular class.
\end{quote}

\newpage

\setcounter{section}{0}
\section{Introduction}
\setcounter{equation}{0}

A Lie algebroid is a natural generalization of the notion
of Lie algebra, and also of the tangent bundle of a
manifold. There are many other interesting examples, for instance, the
cotangent bundle of any Poisson manifold possesses a natural structure
of Lie algebroid. Roughly speaking, a Lie algebroid over a manifold $M$
is a vector bundle $E$ over $M$ such that its space of sections
$\Gamma(E)$ has a structure of Lie algebra plus a mapping (the anchor
map) from $E$ onto $TM$ which provides a Lie algebra homomorphism from
$\Gamma(E)$ into the Lie algebra of vector fields $\hbox{\fr X}(M)$.

The action of $\Gamma(E)$ on $C^{\infty}(M,\R)$ defines the Lie
algebroid cohomology of $M$. For a Poisson manifold $M$, the associated
Lie algebroid is just the triple $(T^{*}M, \lcf \;,\;\rcf, \#),$ where
$\lcf \;,\;\rcf$ is the bracket of 1-forms and $\#$ is the mapping from
1-forms into tangent vectors defined by the Poisson tensor.
For an oriented Poisson manifold $M$ and its associated Lie
algebroid, A. Weinstein \cite{We,We2}   
has defined the so-called modular class of $M$, which is an element of the
corresponding Lie algebroid cohomology (in fact, an element of the 
Lichnerowicz-Poisson cohomology space $H^{1}_{LP}(M)$).
The modular class $X_{\nu}$ is defined as the operator which
assigns to each function $f$ the divergence with respect to $\nu$ of
its Hamiltonian vector field $X_{f}$, where $\nu$ is a volume
form on $M$. A direct computation shows that
the modular class of a symplectic manifold is null. Indeed, the
vanishing of the modular class of a Poisson manifold is closely
related with its regularity. Moreover, it was proved by P. Xu
\cite{Xu} (see also \cite{BZ,EL}) that the canonical homology is dual to the
Lichnerowicz-Poisson cohomology for unimodular Poisson structures,
that is, for Poisson structures with null modular class. Also, it
should be remarked that the modular class was the tool
recently used by J.-P. Dufour and A. Haraki \cite{DH} and by Z.J. Liu
and P. Xu \cite{LX} to classify quadratic Poisson structures. 

Our interest is to extend the above results for Nambu-Poisson
manifolds. The concept of a Nambu-Poisson structure was introduced by
Takhtajan \cite{T} in order to find an axiomatic formalism for the
$n$-bracket operation 
\[
\{f_1,\dots ,f_n\}=\mbox{\rm det}(\frac{\partial f_i}{\partial x_j})
\]
proposed by Nambu \cite{Nam} to generalize Hamiltonian mechanics
(see also \cite{BF,Cha,ChT,FDS}). A Nambu-Poisson manifold is a manifold
$M$ endowed with a skew-symmetric 
$n$-bracket of functions $\{\,, \dots, \,\}$ satisfying
the Leibniz rule and the fundamental identity 
\[
\{f_1, \dots, f_{n-1}, \{g_1, \dots, g_n\} \} =
\sum_{i=1}^{n} \{g_1, \dots, \{f_1, \dots, f_{n-1}, g_i\}, \dots, g_n
\} ,
\]
for all $f_1, \dots, f_{n-1}, g_1, \dots, g_{n} \in C^{\infty}(M,\R)$. 
The local and global structure of a Nambu-Poisson manifold were
ellucidated in recent papers \cite{AlGu,Gau,ILMM,MVV,Nak}. Indeed, a
Nambu-Poisson manifold of order 
greater than 2 consists of pieces which are volume manifolds, in the
same way that a Poisson manifold is made of symplectic pieces.
Recently, an interesting recursive characterization of Nambu-Poisson
structures was obtained in \cite{GM}. 

In this paper we introduce the notion of a Leibniz
algebroid -a natural generalization of a Lie algebroid. 
The notion of Leibniz algebra was recently introduced by J.L. Loday
\cite{L1,L2} (see also \cite{LP}) as a noncommutative version of Lie algebras. Indeed, a
Leibniz algebra is a real vector space $\hbox{\fr g}$ endowed with a
$\R$-bilinear mapping $\{\,,\,\}$ satisfying the Leibniz identity
\[
\{a_1,\{a_2,a_3\}\}-\{\{a_1,a_2\},a_3\}-\{a_2,\{a_1,a_3\}\}=0,  
\]
for all $a_1, a_2, a_3 \in {\frak g}$.
If the bracket is skew-symmetric we recover the notion of Lie algebra.
Next, the notion of Leibniz algebroid can be introduced in the same
way that for the case of Lie algebroids. One of the main results of the
present paper is to associate a Leibniz algebroid to each Nambu-Poisson
manifold $M$. 
The Leibniz algebroid attached to $M$ is just the triple
$(\bigwedge^{n-1}(T^*M),$ $\lcf \;,\;\rcf ,\#)$, where 
$\lcf \;,\;\rcf:\Omega^{n-1}(M) \times \Omega^{n-1}(M) \longrightarrow
\Omega^{n-1}(M)$ is the bracket of $(n-1)$-forms defined by 
\[
\lcf \alpha,\beta \rcf = {\cal L}_{\#\alpha}\beta+
(-1)^n(i(d\alpha)\Lambda)\beta, 
\]
for $\alpha, \beta \in \Omega^{n-1}(M)$, and
$\# : \bigwedge^{n-1}(T^*M) \longrightarrow TM$ is the homomorphism
of vector bundles given by $\#(\beta) = i(\beta)\Lambda$.
Here $\Lambda$ is the Nambu-Poisson $n$-vector.
In addition, it is proved that the only non-null Nambu-Poisson
structures of order greater than 2 on an oriented manifold $M$ of
dimension $m$, with $m\geq 3,$ 
such that its Leibniz algebroid is a Lie algebroid are those defined by
non-null $m$-vectors.

As in the case of Poisson manifolds, we define the modular class of an
oriented Nambu-Poisson $m$-dimensional manifold $M$ of order $n$.
Indeed, if $\nu$ is a volume form on $M$ then 
the mapping 
\[
{\cal M}_{\nu} : C^\infty(M,\R) \times \dots^{(n-1} \dots \times
C^\infty(M,\R) \longrightarrow C^\infty(M,\R),
\]
given by 
\[
{\cal L}_{X_{f_1 \dots f_{n-1}}} \nu = {\cal M}_{\nu}(f_1, \dots,
f_{n-1}){\nu}
\]
is a $(n-1)$-vector on $M$, where $X_{f_1\dots f_{n-1}}=\#(df_1\wedge
\dots \wedge df_{n-1})$ is the Hamiltonian vector field associated
with the functions $f_1,\dots ,f_{n-1}$. Next, the mapping 
\[
{\cal M}_{\Lambda} : \Omega^{n-1}(M) \longrightarrow
C^\infty(M,\R) \makebox[1cm]{} \alpha \mapsto i(\alpha){\cal M}_\nu 
\]
defines a $1$-cocycle in the Leibniz cohomology complex associated to
the Leibniz algebroid $(\mbox{$\bigwedge^{n-1}$}(T^*M),$ $\lcf
\;,\;\rcf, \#)$. The cohomology class $[{\cal M}_{\Lambda}]\in 
H^1(\Omega^{n-1}(M);C^\infty(M,\R))$ does not depend on the chosen
volume form and it is called the modular class of
$M$. As in the case of Poisson manifolds it is proved that the modular
class of a volume manifold is null. In a forthcoming paper we will
investigate the role played for the modular class in the problem of
classification of Nambu-Poisson manifolds. We also are investigating
the existence of a dual homology to the Leibniz algebroid cohomology,
which would be related with the vanishing of the modular class (see
\cite{BZ,EL,Xu} for the case of Lie algebroids and Poisson manifolds).

\section{Preliminaries}
\setcounter{equation}{0}
All the manifolds considered in this paper are assumed to be connected.
\subsection{Nambu-Poisson structures}

Let $M$ be a differentiable manifold of dimension $m$. Denote by
${\frak X}(M)$ the Lie algebra of vector fields on $M$, by
$C^\infty(M,\R)$ the algebra of $C^\infty$ real-valued functions on
$M$ and by $\Omega^{k}(M)$ the space of $k$-forms on $M$. 

{\it An almost Poisson bracket} of order $n$ $(n\leq m)$ on $M$ (see
\cite{ILMM}) is an $n$-linear 
mapping $\{,\dots,\}:C^\infty(M,\R)\times \dots^{(n}\dots  \times
C^\infty(M,\R)\rightarrow C^\infty(M,\R)$ satisfying the following
properties : 
\begin{enumerate}
\item[$(1)$] {\it (Skew-symmetry)}
\begin{equation}\label{skew}
\{f_1,\dots ,f_n\}=(-1)^{\varepsilon(\sigma)}\{f_{\sigma(1)},\dots
,f_{\sigma(n)}\}, 
\end{equation}
for all $f_1,\dots ,f_n\in C^\infty(M,\R)$ and $\sigma\in Symm(n)$,
where $Symm(n)$ is a symmetric group of $n$ elements and
$\varepsilon(\sigma)$ is the parity of the permutation $\sigma$. 
\item[$(2)$] {\it (Leibniz rule)}
\begin{equation}\label{rL}
\{f_1g_1, f_2,\dots ,f_n\}=f_1\{g_1,f_2,\dots ,f_n\}+g_1\{f_1,f_2,\dots
,f_n\},
\end{equation}
for all $f_1,\dots ,f_{n},g_1\in C^\infty(M,\R).$ 
\end{enumerate}

If $\{,\dots ,\}$ is an almost Poisson bracket of order $n$ then 
we define a skew-symmetric tensor $\Lambda$ of type $(n,0)$
($n$-vector) as follows
\[
\Lambda(df_1,\dots ,df_n)=\{f_1,\dots,f_n\},
\]
for $f_1,\dots ,f_n\in C^\infty(M,\R).$ Conversely, given an
$n$-vector on $M$, the above formula defines an almost Poisson
bracket of order $n$. The pair $(M,\Lambda)$ is called  a {\it generalized
almost Poisson manifold of order n}.

If $\bigwedge^{n-1}(T^{*}M)$ is the vector bundle of the $(n-1)$-forms
on $M$ then $\Lambda$ induces a homomorphism of vector bundles
\[
\#: \mbox{$\bigwedge^{n-1}$}(T^{*}M) \longrightarrow TM
\]
by defining
\begin{equation}\label{sos'}
\#(\beta) = i(\beta)\Lambda(x)
\end{equation}
for $\beta \in \bigwedge^{n-1}(T_{x}^{*}M)$ and $x \in M$, where
$i(\beta)$ is the contraction by $\beta$. Denote also by
\[
\#:\Omega^{n-1}(M)\rightarrow {\frak X}(M)
\]
the homomorphism of $C^{\infty}(M, \R)$-modules given by 
\begin{equation}\label{sos}
\#(\alpha)(x) = \#(\alpha(x))
\end{equation}
for all $\alpha \in \Omega^{n-1}(M)$ and $x \in M$.
Then, if $f_1,\dots ,f_{n-1}$ are $n-1$ functions on $M$, we define a
vector field 
\begin{equation}\label{ch}
X_{f_1\dots f_{n-1}}=\#(df_1\wedge \dots \wedge df_{n-1}),
\end{equation}
which is called the {\it Hamiltonian vector field } associated with
the Hamiltonian functions $f_1,\dots,$ $f_{n-1}$. From (\ref{ch}) it
follows that 
\begin{equation}\label{2.5'}
X_{f_1\dots f_{n-1}}(f_n)=\{f_1, \dots, f_{n-1}, f_{n}\}.
\end{equation}

A more rich structure, related with interesting dynamical problems,
can be considered adding to the almost Poisson bracket $\{,\dots ,\}$
the following integrability condition  
({\it fundamental identity}) 
\begin{equation}\label{if}
\{f_1,\dots ,f_{n-1},\{g_1,\dots ,g_n\}\}=
\sum_{i=1}^{n}\{g_1,\dots ,\{f_1,\dots ,f_{n-1},g_i\},\dots ,g_n\}
\end{equation}
for all $f_1,\dots ,f_{n-1},g_1,\dots ,g_{n}$ functions on $M.$ In this case,
$\{,\dots ,\}$  is called a {\it Nambu-Poisson
bracket } and $(M,\Lambda)$ is a {\it Nambu-Poisson manifold of order
$n$ } (see \cite{T}). 

In fact, an $n$-vector $\Lambda$ on $M$ defines a Nambu-Poisson
structure if the Hamiltonian vector fields are derivations of the
algebra $(C^\infty(M,\R)\times \dots^{(n}\dots \times C^\infty(M,\R),
\{,\dots,\})$ or equivalently, every Hamiltonian vector field
$X_{f_1\dots f_{n-1}}$ is an infinitesimal automorphism of $\Lambda$,
that is, 
\[
{\cal L}_{X_{f_1\dots f_{n-1}}}\Lambda=0.
\]

\begin{examples}\label{2.1}{\rm 
(i) The {\it Poisson manifolds } are just the Nambu-Poisson manifolds of order
$2$ \cite{Li,Vais,W}. 

Another examples of Nambu-Poisson manifolds are the following.

(ii) Let $N$ be an oriented $m$-dimensional manifold and
choose a vo\-lu\-me form $\nu_N$ on $N$. Given $m$ functions $f_1,\dots
,f_m$ on $N$, we define its $m$-bracket by the formula 
\begin{equation}\label{vol}
df_1\wedge\dots \wedge df_m=\{f_1,\dots ,f_m\}\nu_N.
\end{equation}
It is not hard to prove that it is a Nambu-Poisson bracket (see
\cite{Gau}). Denote
by $\Lambda_{\nu_N}$ the $m$-vector associated with this bracket.
Note that for the Nambu-Poisson structure $\Lambda_{\nu_{N}}$ the
homomorphism $\#: \Omega^{m-1}(N) \longrightarrow {\frak X}(N)$ is an
isomorphism. Furthermore, if $\Lambda$ is a Nambu-Poisson structure
of order $m$ and $\Lambda \not= 0$ at every point then there exists a
volume form $\nu$ on $N$ such that $\Lambda = \Lambda_{\nu_{N}}$ (see
\cite{ILMM}).

(iii) Let $\Lambda_N$ be an arbitrary $m$-vector on
an oriented $m$-dimensional manifold $N$ with volume form $\nu_N$.
Then, there exists a function $f\in C^\infty(N,\R)$ such that
$\Lambda_N=f\Lambda_{\nu_N}$. Moreover, if $f_{1}, \dots , f_{m-1}$
are $m-1$ functions on $N$ and $X_{f_1\dots f_{m-1}}$ is the
corresponding Hamiltonian vector field with respect to the $m$-vector
$\Lambda_{N}$, it follows that ${\cal L}_{X_{f_1\dots
f_{m-1}}}\Lambda_{N} = 0$. Thus, we deduce that
$(N,\Lambda_N)$  is a Nambu-Poisson manifold  of order $m$. 

(iv) If $V$ is a $k$-dimensional differentiable manifold,
$\Lambda_N$ induces an $m$-vector $\Lambda$  on the product $N\times
V$ and $(N \times V,\Lambda)$ is a Nambu-Poisson manifold of order
$m$. } 

\end{examples}

The following theorem describes the local structure of
the Nambu-Poisson brackets of order $n$, with $n \geq 3$. 

\begin{theorem}\label{thl}\cite{AlGu,Gau,ILMM,MVV,Nak}
Let $M$ be a differentiable manifold of dimension $m$. The $n$-vector
$\Lambda$, $n \geq 3$,  
defines a Nambu-Poisson structure on $M$ if and only if for all $x\in
M$ where $\Lambda(x)\not=0,$ there exist local coordinates $(x^1,\dots
,x^n,x^{n+1},\dots ,x^m)$ around $x$ such that 
\[
\Lambda=\frac{\partial}{\partial x^1}\wedge \dots \wedge
\frac{\partial}{\partial x^n}.
\]
\end{theorem}

\begin{remark}{\rm A point $x$ of a Nambu-Poisson manifold
$(M,\Lambda)$ of order $n \geq 3$ is said to be {\it regular } if $\Lambda
(x) \not= 0$.}
\end{remark}

\begin{remark}{\rm Let $(M,\Lambda)$ be a Nambu-Poisson manifold of
order $n$, with $n\geq 3,$ and consider the {\it characteristic
distribution } ${\cal D}$ on $M$ given by 
\[
x\in M\longrightarrow {\cal
D}(x)=\#(\Lambda^{n-1}T_x^*M)=<\{X_{f_1\dots f_{n-1}}(x)/f_1,
\dots ,f_{n-1}\in C^\infty(M,\R)\}>\subseteq T_xM.
\]
Then, ${\cal D}$ defines a generalization foliation on $M$ whose
leaves are either points or $n$-dimensional manifolds endowed with a
Nambu-Poisson structure coming from a volume form (see \cite{ILMM}).}
\end{remark}

\subsection{Leibniz algebras and cohomology}

First, we recall the definition of real Leibniz algebra (see
\cite{L1,L2,LP}). 

A {\it Leibniz algebra structure}  on a
real vector space ${\frak g}$ is a $\R$-bilinear map $\{\;,\;\}:{\frak
g}\times {\frak g}\rightarrow {\frak g}$ satisfying the {\it Leibniz
identity}, that is, 
\begin{equation}
\{a_1,\{a_2,a_3\}\}-\{\{a_1,a_2\},a_3\}-\{a_2,\{a_1,a_3\}\}=0,
\end{equation}
for $a_1, a_2, a_3\in {\frak g}.$ In such a case, one says that
$({\frak g},\{\;,\;\})$ is a {\it Leibniz algebra}.

Moreover, if the skew-symmetric
condition is required then $({\frak g}, \{\;,\;\})$ is a Lie algebra.
In this sense, a Leibniz algebra is a non-commutative version of a
Lie algebra.

Let  $({\frak g},\{\;,\;\})$ be a Leibniz algebra and ${\cal M}$ be a
real vector space endowed with a $\R$-bilinear map
\[
{\frak g}\times {\cal M}\longrightarrow {\cal M}
\]
such that $\{a_1,a_2\}m=a_1(a_2m)-a_2(a_1m),$ for all $a_1,a_2\in
{\frak g}$ and $m\in {\cal M}.$ Then 
${\cal M}$ is a {\it ${\frak g}$-module
relative to the representation} of ${\frak g}$ on ${\cal M}.$ 

If ${\cal M}$ is a ${\frak g}$-module we can introduce a cohomology
complex as follows.

A $k$-linear mapping $c^k:{\frak g}\times \dots^{(k}\dots\times{\frak g}
\longrightarrow {\cal M}$ is called 
a {\it ${\cal M}$-valued $k$-cochain}. We denote by $C^k({\frak g};{\cal
M})$ the real vector space of these cochains.

The operator $\partial^k:C^{k}({\frak g};{\cal M}) \longrightarrow
C^{k+1}({\frak g}; {\cal M})$ given by 
\begin{equation}\label{cohole}
\begin{array}{lcl}
\partial^kc^k(a_0, \dots,
a_k)&=&\displaystyle\sum_{i=0}^k(-1)^ia_ic^k(a_0,\dots , 
\widehat{a_i}, \dots, a_k)\\ 
&\kern-40pt+&\kern-40pt\displaystyle\sum_{0\leq i<j\leq
k}(-1)^{i-1}c^k(a_0, \dots ,\widehat{a_i},\dots
,a_{j-1},\{a_i,a_j\},a_{j+1},\dots ,a_k), 
\end{array}
\end{equation}
defines a coboundary since $\partial^{k+1}\circ \partial^k=0.$ Hence,
$(C^*({\frak g};{\cal M}),\partial)$ is a cohomology complex and  
the corresponding cohomology spaces
\[
H^k({\frak g};{\cal M})=\frac{\mbox{\rm ker}\{\partial^k:C^k({\frak
g};{\cal M})\rightarrow C^{k+1}({\frak g};{\cal M})\}}{\mbox{\rm
Im}\{\partial^{k-1}:C^{k-1}({\frak g}; {\cal M})\rightarrow
C^k({\frak g}; {\cal M})\}},
\]
are called the {\it Leibniz cohomology groups of ${\frak g}$ with
coefficients in ${\cal M}$} (see \cite{L1,L2,LP}). 

Note that if $({\frak g},\{\;,\;\})$ is a Lie algebra and $c^k$ is a
skew-symmetric ${\cal M}$-valued $k$-cochain then $\partial^k c^k$ is
a skew-symmetric ${\cal M}$-valued $(k+1)$-cochain. Thus, we can
consider the subcomplex $(C^*_{Lie}({\frak g}; {\cal M}), \partial)$
of $(C^*({\frak g}; {\cal M}), \partial)$ that consists of the 
skew-symmetric ${\cal M}$-valued cochains. In fact, the cohomology of
this subcomplex is just the cohomology $H_{Lie}^*({\frak
g};{\cal M})$ of the Lie algebra ${\frak g}$ with coefficients in
${\cal M}$. Therefore, we have defined a natural homomorphism
\[
i^{k}: H^k_{Lie}({\frak g}; {\cal M})\rightarrow H^k({\frak g};{\cal
M})
\]
between the cohomology groups $H^k_{Lie}({\frak g}; {\cal M})$ and
$H^k({\frak g};{\cal M})$.

\begin{examples}{\rm 
(i) Let $M$ be a differentiable manifold and $({\frak X}(M),[\;,\;])$
the Lie algebra of the vector fields on $M$. 
Then, the real vector space $C^\infty(M,\R)$ is a ${\frak
X}(M)$-module with the usual multiplication 
\[
{\frak X}(M)\times C^\infty(M,\R)\longrightarrow
C^\infty(M,\R),\makebox[1cm]{} (X,f)\mapsto X(f).
\]
The $k$-cochains in the Leibniz cohomology complex are
the $k$-linear mappings $c^k:{\frak X}(M)\times \dots^{(k}\dots\times
{\frak X}(M)\rightarrow C^\infty(M,\R)$ and the Leibniz cohomology operator
$d:C^k_{Leib}(M) = \linebreak C^k({\frak X}(M);C^\infty(M,\R))\longrightarrow
C^{k+1}_{Leib}(M) = C^{k+1}({\frak X}(M);C^\infty(M,\R))$
is defined as the exterior differential operator, that is, 
\begin{equation}\label{d}
\begin{array}{lcl}
dc^k(X_0,\dots ,X_k)&=&\displaystyle\sum_{i=0}^k(-1)^iX_i(c^k(X_0,\dots
,\widehat{X_i}, \dots, X_k))\\
&\kern-40pt+&\kern-40pt\displaystyle\sum_{0\leq i<j\leq
k}(-1)^{i-1}c^k(X_0, 
\dots ,\widehat{X_i},\dots ,X_{j-1},[X_i,X_j],X_{j+1},\dots ,X_k),
\end{array}
\end{equation}
for all $X_0,\dots ,X_k\in {\frak X}(M).$

The resultant cohomology $H^*({\frak X}(M);C^\infty(M,\R))$ is the
{\it Leibniz cohomology } of $M$ and it is denoted by $H^*_{Leib}(M)$ (for a
detailed study of this cohomology, we refer to \cite{Lo}). Note that
the de Rham cohomology of $M$, $H^{*}_{dR}(M)$, is just the
cohomology of the subcomplex of the skew-symmetric $C^{\infty}(M,
\R)$-valued cochains that are $C^{\infty}(M, \R)$-linear. 

(ii) Let $(M,\Lambda)$ be a Nambu-Poisson manifold
of order $n$ with Nambu-Poisson bracket $\{,\dots ,\}$. 
Consider on $\bigwedge^{n-1}(C^\infty(M,\R))$ the bracket $\{\; ,
\;\}'$ characterized by the formula:
\begin{equation}\label{cF}
\{f_1\wedge \dots \wedge f_{n-1},g_1\wedge \dots \wedge g_{n-1}\}'=
\sum_{i=1}^{n-1}g_1\wedge \dots \wedge \{f_1,\dots ,f_{n-1},g_i\}\wedge
\dots \wedge g_{n-1}
\end{equation}
for $f_1,\dots ,f_{n-1},g_1,\dots ,g_{n-1}\in C^\infty(M,\R).$
Using (\ref{if}), we deduce that
$(\bigwedge^{n-1}(C^\infty(M,\R)), \linebreak \{\;,\;\}')$ is a Leibniz 
algebra (see \cite{DT,Gau,Nak}). Moreover, the real vector space 
$C^{\infty}(M, \R)$ is a $\bigwedge^{n-1}(C^\infty(M,\R))$-module
with the multiplication 
\[
\mbox{$\bigwedge^{n-1}(C^\infty(M,\R))$}\times C^\infty(M,\R)\longrightarrow
C^\infty(M,\R)
\]
characterized by
\begin{equation}\label{rF}
(f_1\wedge \dots \wedge f_{n-1},f)\mapsto X_{f_1\dots f_{n-1}}(f).
\end{equation}
The resultant cohomology
\[
H^*(\mbox{$\bigwedge^{n-1}$}(C^\infty(M,\R));C^\infty(M,\R))
\] 
was studied in \cite{DT} and \cite{Gau}.  
}
\end{examples}

\section{Leibniz algebroids and Nambu-Poisson  manifolds}
\setcounter{equation}{0}

In this section we will define a generalization of the notion of Lie
algebroid (the Leibniz al\-ge\-broid) and we will prove that a
Nambu-Poisson manifold has associated a structure of this type.

\begin{definition}\label{alg}
A Leibniz algebroid structure on a differentiable vector bundle
$\pi:E\rightarrow M$ is a pair that consists of a Leibniz algebra
structure $\lcf \;,\; \rcf $ on the space $\Gamma(E)$ of the global cross
sections of $\pi:E\longrightarrow M$ and a vector bundle morphism 
$\varrho:E\rightarrow TM,$ called the anchor map, such that the
induced map $\varrho:\Gamma(E)\longrightarrow \Gamma(TM)={\frak
X}(M)$ satisfies the following relations:
\begin{enumerate}
\item
$\varrho\lcf s_1,s_2 \rcf =[\varrho(s_1),\varrho(s_2)],$
\item
$\lcf s_1,fs_2\rcf =f\lcf s_1,s_2\rcf +\varrho(s_1)(f)s_2,$
\end{enumerate}
for all $s_1,s_2\in \Gamma(E)$ and $f\in C^\infty(M,\R).$ 

A triple $(E,\lcf \;,\; \rcf,\varrho)$ is called a Leibniz algebroid
over $M$. 
\end{definition}

\begin{remark} {\rm
\begin{enumerate}
\item
If $({\frak g},\{\;,\;\})$ is a Leibniz algebra then $({\frak
g},\{\;,\;\},\varrho\equiv 0)$ is a Leibniz algebroid over a point.
\item
Every Lie algebroid over a manifold $M$ is trivially a Leibniz algebroid. 
In fact, a Leibniz algebroid $(E,\lcf \;,\; \rcf,\varrho)$ over $M$ is a
Lie algebroid if and only if the Leibniz bracket $\lcf \;,\; \rcf$ on
$\Gamma(E)$ is skew-symmetric.
\end{enumerate}}
\end{remark}

If $(M,\Lambda)$ is a Poisson manifold then it is possible to define
a Lie algebra structure $\lcf \;,\; \rcf$ on the space of $1$-forms
$\Omega^1(M)$ in such a sense that the triple $(T^*M,\lcf \;,\; \rcf,\#)$
is a Lie algebroid over $M$, where $T^*M$ is the cotangent bundle of
$M$ and $\#:T^*M \rightarrow TM$ is the homomorphism of vector bundles
given by (\ref{sos}) (see \cite{BV,Vais}).

Next, we will prove that associated to a Nambu-Poisson manifold of
order $n$, with $n\geq 3$, there is a canonical Leibniz algebroid. 

Let $(M,\Lambda)$ be an $m$-dimensional Nambu-Poisson manifold of
order $n, n\geq 3$, with Nambu-Poisson bracket $\{\;,\dots,\;\}.$

\begin{proposition}\label{4.2}
For all $\alpha,\beta\in \Omega^{n-1}(M)$ we have
\[
[\#\alpha,\#\beta]=\#({\cal L}_{\#\alpha}\beta+(-1)^n(i(d\alpha)\Lambda)\beta)
\]
where $\# : \Omega^{n-1}(M) \longrightarrow {\frak X}(M)$ is the
homomorphism defined in (\ref{sos'}) and (\ref{sos}) and ${\cal L}$ is
the Lie derivative operator. 
\end{proposition}
{\bf Proof:} Using (\ref{sos'}) and (\ref{sos}) we have that 
\begin{equation}\label{CX}
\begin{array}{lcl}
[\#\alpha,\#\beta]&=&
i(\beta)({\cal
L}_{\#\alpha}\Lambda)+ i({\cal
L}_{\#\alpha}\beta)\Lambda\\
&=&i(\beta)({\cal L}_{\#\alpha}\Lambda)+ \#({\cal L}_{\#\alpha}\beta).
\end{array}
\end{equation}
 
On the other hand, 
\begin{equation}\label{X}
{\cal L}_{\#\alpha}\Lambda=(-1)^n(i(d\alpha)\Lambda)\Lambda.
\end{equation}
Indeed, if $x\in M$ and $\Lambda(x)=0$ then $({\cal
L}_{\#\alpha}\Lambda)(x)=(-1)^n(i(d\alpha)\Lambda)(x)\Lambda(x)=0.$

If $x\in M$ and $\Lambda(x)\not=0,$ then (see Theorem \ref{thl}) there
exist local coordinates $(x^1,\dots ,x^n,$ $x^{n+1},\dots, x^m)$ in an open
subset $U$ of $M$, $x\in U,$ such that 
\begin{equation}\label{lnp}
\Lambda=\frac{\partial}{\partial x^1}\wedge \dots \wedge
\frac{\partial }{\partial x^n}.
\end{equation}
So, to prove (\ref{X}) it suffices of course to check this formula for
local $(n-1)$-forms
\[
\alpha=\sum_{i=1}^n\alpha_idx_1\wedge \dots \widehat{dx_i}\wedge
\dots \wedge dx_n,
\]
with $\alpha_i\in C^\infty(U,\R).$ Now, from (\ref{sos'}), (\ref{sos}) and
(\ref{lnp}), one deduces that (\ref{X}) is true for this type of $(n-1)$-forms.

Finally,  using (\ref{sos'}), (\ref{sos}), (\ref{CX}) and
(\ref{X}), the result follows. 
\hfill$\Box$

The above result suggests us to  introduce the following definition.

\begin{definition}\label{3.3'}
Let $(M,\Lambda)$ be an $m$-dimensional Nambu-Poisson manifold of
order $n,$ with $3\leq n\leq m.$ The bracket of $(n-1)$-forms on $M$
is the $\R$-bilinear operation  
$\lcf \;,\; \rcf:\Omega^{n-1}(M)\times
\Omega^{n-1}(M)\longrightarrow \Omega^{n-1}(M)$ given by
\begin{equation}\label{c1f}
\lcf \alpha,\beta \rcf ={\cal L}_{\#\alpha}\beta+
(-1)^n(i(d\alpha)\Lambda)\beta, 
\end{equation}
for $\alpha,\beta\in \Omega^{n-1}(M)$. 
\end{definition}

The mapping $\lcf \;,\; \rcf$ is characterized as follows:

\begin{theorem}
Let $(M,\Lambda)$ be an $m$-dimensional Nambu-Poisson
manifold of order $n,$ with $3\leq n\leq m$. Then exists a unique
$\R$-bilinear operation $\lcf \;,\; \rcf :\Omega^{n-1}(M)\times
\Omega^{n-1}(M)\rightarrow \Omega^{n-1}(M)$ such that: 
\begin{enumerate}
\item For all $f_1,\dots ,f_{n-1},g_1,\dots ,g_{n-1}\in
C^\infty(M,\R)$, we have
\begin{equation}\label{c1}
\kern-30pt\lcf df_1\wedge \dots \wedge df_{n-1},dg_1\wedge \dots \wedge
dg_{n-1}\rcf =\displaystyle\sum_{i=1}^{n-1} dg_1\wedge \dots \wedge
d\{f_1,\dots ,f_{n-1},g_i\}\wedge 
\dots \wedge dg_{n-1}.
\end{equation}
\item For all $f\in C^\infty(M,\R)$ and $\alpha,\beta\in
\Omega^{n-1}(M)$, we have 
\begin{equation}\label{c21}
\lcf \alpha,f\beta\rcf =f\lcf \alpha,\beta\rcf +\#\alpha(f)\beta,
\end{equation}
\begin{equation}\label{c22}
\lcf f\alpha,\beta\rcf =f\lcf \alpha,\beta\rcf -i(\#\alpha)(df\wedge
\beta).
\end{equation}
\end{enumerate}
This operation is given by (\ref{c1f}).
\end{theorem}

{\bf Proof:}
It is easy to prove that the bracket defined in (\ref{c1f}) satisfies
(\ref{c1}), (\ref{c21})  and (\ref{c22}).

Now, suppose that  $\lcf \;,\;\rcf_1:\Omega^{n-1}(M)\times
\Omega^{n-1}(M)\longrightarrow \Omega^{n-1}(M)$ is
a $\R$-bilinear operation which satisfies (\ref{c1}), (\ref{c21}) and
(\ref{c22}).  Then $\lcf \;,\;\rcf_1 $ must be of the local type, i.e.,
$\lcf \alpha,\beta\rcf_1(x_0)$ will depend on $\alpha$ and $\beta$ around
$x_0$ only, for all $x_0\in M.$ Indeed, if $\beta_{1|U}=\beta_{2|U}$
for an open neighborhood $U$ of $x_0$, and if $f$ is a
$C^\infty$ real-valued function that vanishes outside $U$, and equals 1 on a
compact neighborhood $V_{x_0}\subseteq U$, then
$\sigma=f\beta_1=f\beta_2$ is well defined on $M$ and,
by $(\ref{c21})$ we have
\[
\lcf \alpha,\sigma\rcf_1 (x_0)=\lcf \alpha,\beta_1\rcf_1 (x_0),\;\;
\lcf \alpha,\sigma\rcf_1 (x_0)=\lcf \alpha,\beta_2\rcf_1 (x_0),\;\;
\]
i.e., $\lcf \alpha,\beta_1\rcf_1 (x_0)=\lcf \alpha,\beta_2\rcf_1
(x_0)$.  

Similarly, if $\alpha_{1|U}=\alpha_{2|U}$ and
$\nu=f\alpha_1=f\alpha_2$ then, from (\ref{c22}), we deduce that 
\[
\lcf \nu,\beta\rcf_1 (x_0)=\lcf \alpha_1,\beta\rcf_1 (x_0),\;\;
\lcf \nu,\beta\rcf_1 (x_0)=\lcf \alpha_2,\beta\rcf_1 (x_0),\;\;
\]
that is, $\lcf \alpha_1,\beta\rcf_1 (x_0)=\lcf \alpha_2,\beta\rcf_1 (x_0)$.

Next, we will show that $\lcf \;,\;\rcf_1=\lcf \;,\;\rcf $.

Let $x$ be a point of $M$ and $\alpha$ and $\beta$ $(n-1)$-forms on $M$.

Assume that $(x^1,\dots  ,x^m)$ are local coordinates in an open
neighborhood $U$ of $x$ and that in $U$ we have   
\[
\begin{array}{l}
\alpha=\displaystyle\sum_{1\leq i_1< \dots <i_{n-1}\leq m} \alpha_{i_1\dots
i_{n-1}}dx^{i_1}\wedge \dots 
\wedge dx^{i_{n-1}}\\
\beta=\displaystyle\sum_{1\leq j_1<\dots <j_{n-1}\leq m}\beta_{j_1\dots
j_{n-1}} dx^{j_1}\wedge 
\dots \wedge dx^{j_{n-1}}.
\end{array}
\]
Using (\ref{ch}), (\ref{c1f}), (\ref{c1}), (\ref{c21}), (\ref{c22})
and the local character of the bracket $\lcf \;,\;\rcf_1 $, we obtain that   
\[
\begin{array}{lcl}
\lcf \alpha,\beta\rcf_1(x)&=&\kern-40pt \displaystyle\sum_{
\scriptstyle\begin{array}{l}
\scriptstyle 1\leq i_1<\dots <i_{n-1}\leq m\\[-5pt]
\scriptstyle 1\leq j_1<\dots <j_{n-1}\leq m
\end{array}}[\sum_{k=1,\dots ,n-1}(
\alpha_{i_1\dots i_{n-1}}\beta_{j_1\dots j_{n-1}}dx^{j_1}\wedge
\dots \wedge dx^{j_{k-1}}\wedge\\
&&\wedge d\{x^{i_1},\dots, x^{i_{n-1}},x^{j_k}\}\wedge
dx^{j_{k+1}}\dots \wedge dx^{j_{n-1}})\\
&&-\beta_{j_1\dots j_{n-1}}X_{x^{i_1}\dots
x^{i_{n-1}}}(\alpha_{i_1\dots i_{n-1}})dx^{j_1}\wedge \dots \wedge
dx^{j_{n-1}}\\
&&+ \beta_{j_1\dots j_{n-1}} d\alpha_{i_1\dots i_{n-1}}\wedge
i_{X_{x^{i_1}\dots x^{i_{n-1}}}}(dx^{j_1}\wedge 
\dots \wedge dx^{j_{n-1}})\\
&& + \alpha_{i_1\dots i_{n-1}} X_{x^{i_1}\dots
x^{i_{n-1}}}(\beta_{j_1\dots j_{n-1}})dx^{j_1}\wedge
\dots \wedge dx^{j_{n-1}}] (x)\\
&=&\lcf \alpha,\beta\rcf (x).
\end{array}
\]
From the arbitrariness of the point $x$, it follows that 
$\lcf \alpha,\beta\rcf_1=\lcf \alpha,\beta\rcf $. 
\hfill$\Box$

Now, we will prove that a Nambu-Poisson manifold of order $n$, with
$n\geq 3,$ has associated a Leibniz algebroid. 

\begin{theorem}\label{t3.5}
Let $(M,\Lambda)$ be an $m$-dimensional Nambu-Poisson manifold of
order $n$, with $3\leq n\leq m$. Then, the triple
$(\bigwedge^{n-1}(T^*M),$ $\lcf \;,\;\rcf ,\#)$ is a Leibniz 
algebroid over $M$, where $\lcf \;,\;\rcf:\Omega^{n-1}(M)\times
\Omega^{n-1}(M)\longrightarrow \Omega^{n-1}(M)$ is the bracket of
$(n-1)$-forms defined by (\ref{c1f}) and 
$\#:\bigwedge^{n-1}(T^*M)\longrightarrow TM$ is the homomorphism
of vector bundles given by (\ref{sos'}).
\end{theorem}
{\bf Proof:} We must prove that  
\begin{equation}\label{ilcf}
\lcf \alpha,\lcf \beta,\gamma\rcf\rcf -\lcf \lcf
\alpha,\beta\rcf,\gamma\rcf - \lcf \beta,\lcf \alpha,\gamma\rcf\rcf =0,
\end{equation}
for $\alpha, \beta, \gamma \in \Omega^{n-1}(M)$. 

From (\ref{X}), (\ref{c1f})  and Proposition \ref{4.2}, we obtain
that 
\[
(i(d\lcf
\alpha,\beta\rcf)\Lambda-\#\alpha(i(d\beta)\Lambda)+
\#\beta(i(d\alpha)\Lambda)\Lambda=0.
\]
Thus, 
\begin{equation}\label{3.8'}
i(d\lcf
\alpha,\beta\rcf)\Lambda=\#\alpha(i(d\beta)\Lambda)-
\#\beta(i(d\alpha)\Lambda).
\end{equation}
On the other hand, using (\ref{c1f}), we have that 
\[
\lcf\alpha,\lcf\beta,\gamma\rcf\rcf-\lcf\beta,\lcf\alpha,\gamma\rcf\rcf=
{\cal
L}_{[\#\alpha,\#\beta]}\gamma+(-1)^n(\#\alpha(i(d\beta)\Lambda)-
\#\beta(i(d\alpha)\Lambda))\gamma,
\]
which implies that (see (\ref{3.8'}) and Proposition \ref{4.2})
\begin{equation}\label{3.9}
\lcf\alpha,\lcf\beta,\gamma\rcf\rcf-\lcf\beta,\lcf\alpha,\gamma\rcf\rcf=
{\cal
L}_{\#\lcf\alpha,\beta\rcf}\gamma+
(-1)^n(i(d\lcf\alpha,\beta\rcf)\Lambda)\gamma.
\end{equation} 
Therefore, from (\ref{c1f}) and (\ref{3.9}), it follows that
(\ref{ilcf}) holds. Hence, we deduce that the bracket $\lcf\;,\;\rcf$
induces a Leibniz algebra structure on $\Omega^{n-1}(M)$.

Using this fact, (\ref{c21}) and Proposition \ref{4.2},
we conclude that the triple 
$(\bigwedge^{n-1}(T^*M),\linebreak \lcf \;,\;\rcf,\#)$ is a Leibniz
algebroid over $M$. 
\hfill$\Box$

\begin{remark}
{\rm In \cite{GM1} the authors have introduced the notion of Filippov
algebroid, as a $n$-ary generalization of Lie algebroids. Indeed, the
binary bracket of sections in a Lie algebroid is replaced in a Filippov
algebroid $E \longrightarrow M$ by an $n$-bracket $\lcf \;, \dots , \; \rcf$
on $\Gamma(E)$ satisfying the fundamental identity and, the anchor map is a vector
bundle morphism $\Lambda^{n-1}(E) \longrightarrow TM$ compatible with
the $n$-bracket. In \cite{GM1,Vais1} an $n$-bracket of 1-forms on a
Nambu-Poisson manifold is defined. However, this bracket does not
satisfy the fundamental identity. }
\end{remark}

In general, the bracket defined in (\ref{c1f}) is not skew-symmetric
and consequently the Leibniz algebroid
$(\bigwedge^{n-1}(T^*M),\lcf \;,\; \rcf,\#)$ is not a Lie algebroid. 

In the following result we characterize when 
this Leibniz algebroid is a Lie algebroid on an oriented manifold.

\begin{theorem}\label{3.5}
Let $M$ be an oriented manifold of dimension $m$, $m\geq 3.$ The
unique non-null 
Nambu-Poisson structures of order greater than $2$ on $M$ such that
the Leibniz algebroid 
is a Lie algebroid are those defined by 
non-null $m$-vectors.
\end{theorem}
{\bf Proof:} Suppose that $\Lambda$  is a non-null $m$-vector. Then
$(M,\Lambda)$  is a Nambu-Poisson manifold of order $m$ (see Examples
\ref{2.1}).

Now, if $\alpha$ and $\beta$ are $(m-1)$-forms on $M$, we consider the
$(m-1)$-form $\sigma$ on $M$ defined by 
\[
\sigma=\lcf \alpha,\beta\rcf +\lcf \beta,\alpha\rcf.
\]
We must prove that 
$\sigma=0.$

Since the set 
\[
A=\{x\in M/\Lambda(x)\not=0\}
\]
is an open subset of $M$, $\Lambda$ induces a Nambu-Poisson structure
$\Lambda_A$ on $A$ of order $m$ which is  non-null at every point. Then, as $M$ is
oriented, we deduce that $\Lambda_A$ is defined by a volume form on
$A$ and the corresponding homomorphism 
\[
\#_A:\Omega^{n-1}(A)\longrightarrow {\frak X}(A),
\]
given by (\ref{sos'}) and (\ref{sos}), is an isomorphism.
Using this last fact, Proposition \ref{4.2} and the skew-symmetry of
the Lie bracket $[\;,\;]$ of vector fields, we obtain that $\sigma=0$
on $A$.  

On the other hand, it is obvious that  $\sigma$ is null  
on the exterior of $A$ (note that the exterior of $A$ is an open
subset of $M$ and that $\Lambda=0$ on such a set). Finally,  by
continuity we conclude that $\sigma=0$ on the boundary of $A$. Thus,
$\sigma=0$ on $M$.  

\medskip

Conversely, suppose that $(M,\Lambda)$ is an oriented $m$-dimensional
Nambu-Poisson manifold of order $n,$ with $3\leq n\leq m$ and that
$(\bigwedge^{n-1}(T^*M),\lcf \;,\;\rcf,\#)$ is a Lie algebroid.

Since $\Lambda$ is a non-null $n$-vector, there exists a point of $M$
such that $\Lambda(x)\not=0$ and there exist local coordinates
$(x^1,\dots ,x^n,x^{n+1},\dots, x^m)$ on an open neighborhood $U$ of
$x$ such that the $n$-vector $\Lambda_U$ induced by $\Lambda$ on $U$
is given by (see Theorem \ref{thl}) 
\[
\Lambda_U=\frac{\partial }{\partial x^1}\wedge \dots \wedge
\frac{\partial}{\partial x^n}.
\]
Using  the fact that $(\bigwedge^{n-1}(T^*M),\lcf \;,\; \rcf,\#)$
is a Lie algebroid, we deduce that the bracket
$\lcf \;,\; \rcf_U:\Omega^{n-1}(U)\times \Omega^{n-1}(U)\longrightarrow
\Omega^{n-1}(U)$ defined by
$\Lambda_U$ (see (\ref{c1f})) is skew-symmetric.

Now, if $n<m$ we can consider the $n-1$ forms on $U$ given by 
\[
\begin{array}{l}
\alpha=dx^1\wedge \dots \wedge dx^{n-3}\wedge dx^n\wedge dx^{n+1},\\
\beta=x^{n-2}dx^1\wedge \dots \wedge dx^{n-1},
\end{array}
\]
and a direct computation proves that
$0=\lcf \alpha,\beta\rcf_{U}\not=-\lcf \beta,\alpha\rcf_U$. This  is a 
contradiction. Hence, $n=m$. 
\hfill$\Box$

From Theorem \ref{3.5}, we obtain:

\begin{corollary}\label{3.7}
Let $M$ be an oriented manifold of dimension $m$, $m\geq 3$, and
$\nu$ be a volume form on $M$. Then the Leibniz algebroid associated
with the Nambu-Poisson manifold $(M,\Lambda_\nu)$ is a Lie algebroid.
\end{corollary}

\section{Cohomology of a Leibniz algebroid and modular class of a
Nambu-Poisson manifold}
\setcounter{equation}{0}

Let $(E,\lcf \;,\; \rcf,\varrho)$ be a Leibniz algebroid over a manifold
$M.$ From Definition \ref{alg}, we deduce that
$C^\infty(M,\R)$ is a $\Gamma(E)$-module with the multiplication 
\begin{equation}\label{Y}
\Gamma(E)\times C^\infty(M,\R)\longrightarrow
C^\infty(M,\R)\makebox[1cm]{} (s,f)\mapsto \varrho(s)(f).
\end{equation}
Thus, we can consider the differential complex
$(C^*(\Gamma(E);C^\infty(M,\R)),\partial)$ and its cohomology
$H^*(\Gamma(E);C^\infty(M,\R)),$ that is, the cohomology of
$\Gamma(E)$ with coefficients in $C^\infty(M,\R)$ (see Section 2.2).  
$H^*(\Gamma(E);C^\infty(M,\R))$ is called {\it the Leibniz algebroid
cohomology of $E$}. Using (\ref{cohole}), we have that 
\begin{equation}\label{oca}
\begin{array}{lcl}
\partial^kc^k(s_0,\dots , s_k) & = &
\displaystyle\sum_{i=0}^k(-1)^i\varrho(s_i)(c^k(s_0,\dots 
,\widehat{s_i},\dots ,s_k))  \\
&+& \displaystyle\sum_{0\leq i<j\leq
k}(-1)^{i-1}c^k(s_0,\dots,\widehat{s_i},\dots,
s_{j-1},\lcf s_i,s_j\rcf,s_{j+1},\dots ,s_k) 
\end{array}
\end{equation}
for $c^k\in C^k(\Gamma(E);C^\infty(M,\R))$ and $s_0,\dots ,s_k \in
\Gamma(E)$.  

\begin{remark}{\rm 
\begin{enumerate}
\item
Let $(E,\lcf \;,\;\rcf,\varrho)$ be a Lie algebroid over $M$ and
$c^k\in C^k(\Gamma(E);\linebreak C^\infty(M,\R))$.
If $c^k$ is skew-symmetric and $C^\infty(M,\R)$-linear then 
$\partial^kc^k$ is also skew-symmetric and 
$C^\infty(M,\R)$-linear.

The {\it Lie algebroid cohomology } of $E$ is the cohomology of the
subcomplex of the cochains which are skew-symmetric and
$C^\infty(M,\R)$-linear (see \cite{Mac}).

\item
If $(E,\lcf \;,\;\rcf,\varrho)$ is a Leibniz algebroid over $M$ and
$c^k\in C^k(\Gamma(E);C^\infty(M,\R))$ is skew-symmetric
(respectively, $C^\infty(M,\R)$-linear) then, in general,
$\partial^kc^k$ is not skew-symmetric (respectively,
$C^\infty(M,\R)$-linear). 
\end{enumerate}}
\end{remark} 

The following result relates the Leibniz algebroid cohomology of 
$E$ with the Leibniz cohomology of the base manifold $M$.

\begin{proposition}\label{4.1}
Let $(E,\lcf \;,\;\rcf,\varrho)$ be a Leibniz algebroid over a manifold
$M$. Suppose that  $(C_{Leib}^*(M),d)$ is the
Leibniz cohomology complex of the manifold $M$ and denote by 
\[
\tilde{\varrho}^k:C^k_{Leib}(M)\longrightarrow
C^k(\Gamma(E);C^\infty(M,\R)) 
\]
the homomorphism defined by 
\[
\tilde{\varrho}^k(c^k)(s_1,\dots ,s_k)=c^k(\varrho(s_1),\dots,
\varrho(s_k))
\]
for $c^k\in C_{Leib}^k(M)$ and $s_1,\dots ,s_k\in \Gamma(E).$ Then,
the mappings $\tilde{\varrho}^k$ induce a homomorphism of complexes 
\[
\tilde{\varrho}:(C^*_{Leib}(M),d)\longrightarrow
(C^*(\Gamma(E);C^\infty(M,\R)),\partial).
\]
Therefore, we have the corresponding homomorphism in cohomology
\[
\tilde{\varrho}:H^*_{Leib}(M)\longrightarrow
H^*(\Gamma(E);C^\infty(M,\R)). 
\]
\end{proposition}
{\bf Proof:} It follows using (\ref{d}), (\ref{oca}) and
Definition \ref{alg}. 
\hfill$\Box$

\begin{remark}{\rm  
In fact, if $(E,\lcf \;,\;\rcf,\varrho)$ is a Lie
algebroid over $M$, we can define a homomorphism $\bar{\varrho}$
between the de Rham cohomology of $M$, $H_{dR}^*(M)$, and the
Leibniz algebroid cohomology of $E$ given by 
\[
\bar{\varrho} = \tilde{\varrho}\circ i:H^*_{dR}(M)\longrightarrow
H^*_{Leib}(M)\longrightarrow H^*(\Gamma(E);C^\infty(M,\R)),
\]
where $i:H^*_{dR}(M)\longrightarrow H^*_{Leib}(M)$ is
the homomorphism induced by the natural inclusion.}
\end{remark}

Using Proposition \ref{4.1}, we have

\begin{corollary}\label{corol1}
Let $(M,\Lambda)$ be a Nambu-Poisson manifold of order $n,$ with
$n\geq 3,$ and let
$(\bigwedge^{n-1}(T^*M),$ $\lcf \;,\; \rcf,\#)$ be the Leibniz algebroid
associated with $M.$ Suppose that 
\[
\tilde{\#}^k:C^k_{Leib}(M)\longrightarrow
C^k(\Omega^{n-1}(M);C^\infty(M,\R))
\]
is the homomorphism defined by 
\[
\tilde\#^k(c^k)(\alpha_1,\dots
,\alpha_k)=c^k(\#\alpha_1,\dots ,\#\alpha_k),
\]
for $c^k\in C_{Leib}^k(M)$ and $\alpha_1,\dots,
\alpha_k\in \Omega^{n-1}(M).$ Then, the mappings $\tilde{\#}^k$ induce
a homomorphism of complexes
\[
\tilde{\#}:(C_{Leib}^*(M),d)\longrightarrow
(C^*(\Omega^{n-1}(M);C^\infty(M,\R)), \partial ).
\]
Therefore, we have the corresponding homomorphism in cohomology
\[
\tilde{\#}: H_{Leib}^*(M)\longrightarrow H^*(\Omega^{n-1}(M);C^\infty(M,\R)).
\]
\end{corollary}

For the particular case of a Nambu-Poisson structure coming from a
volume form, we deduce the following result:

\begin{proposition}
Let $M$ be an oriented manifold of dimension $m$, with $m\geq 3$,
and let $\nu$ be a volume form on $M$. Then the Leibniz cohomology of
the algebroid associated with $(M,\Lambda_\nu)$ is isomorphic to the
Leibniz cohomology of $M$. 
\end{proposition} 
{\bf Proof:} Since $\nu$ is a volume form, the homomorphism 
\[
\#:\Omega^{m-1}(M)\longrightarrow {\frak X}(M),
\]
defined by (\ref{sos'}) and (\ref{sos}), is an isomorphism.

Using this fact and Corollary \ref{corol1} the result follows.
\hfill$\Box$ 

\begin{remark}{\rm 
Note that the Leibniz algebroid
$(\bigwedge^{n-1}(T^*M),\lcf \;,\;\rcf,\#)$ associated with
$(M,\Lambda_\nu)$ is also a Lie algebroid (see Corollary \ref{3.7})
and that the Lie algebroid cohomology of $\bigwedge^{n-1}(T^*M)$ is
isomorphic to the de Rham cohomology of $M.$ }
\end{remark}

For a Nambu-Poisson manifold $(M, \Lambda)$ of order $n$, we denote
by $\{\;,\;\}'$ the Leibniz bracket on $\bigwedge^{n-1}(C^\infty(M,\R))$
characterized by (\ref{cF}). Then, the real vector space
$C^{\infty}(M, \R)$ is a $\bigwedge^{n-1}(C^\infty(M,\R))$-module
with the multiplication given by (\ref{rF}). Thus, we can consider
the corresponding differential complex
$(C^*(\bigwedge^{n-1}(C^\infty(M,\R));C^\infty(M,\R)),\partial')$ 
and its cohomology
$H^*(\bigwedge^{n-1}(C^\infty(M,\R));C^\infty(M,\R)).$ 

In the next result, we obtain a relation between the Leibniz
algebroid cohomology of $\bigwedge^{n-1}(T^*M)$ and the cohomology
$H^*(\bigwedge^{n-1}(C^\infty(M,\R));C^\infty(M,\R))$. 

\begin{proposition}
Let $(M,\Lambda)$ be a Nambu-Poisson manifold of order $n$, with
$n\geq 3.$ Then the
mapping 
\begin{equation}\label{phi}
\Phi:\mbox{$\bigwedge^{n-1}$}(C^\infty(M,\R))\longrightarrow
\Omega^{n-1}(M),\makebox[1cm]{}f_1\wedge \dots \wedge f_{n-1}\mapsto
df_1\wedge \dots \wedge df_{n-1}
\end{equation}
induces a natural homomorphism of complexes 
\[
\tilde\Phi:(C^*(\Omega^{n-1}(M);C^\infty(M,\R)),\partial)\longrightarrow
(C^*(\mbox{$\bigwedge^{n-1}$}(C^\infty(M,\R));C^\infty(M,\R)),\partial')
\]
and therefore we have the corresponding homomorphism in cohomology 
\[
\tilde\Phi:H^*(\Omega^{n-1}(M);C^\infty(M,\R))\longrightarrow
H^*(\mbox{$\bigwedge^{n-1}$}(C^\infty(M,\R));C^\infty(M,\R)).
\]
\end{proposition}
{\bf Proof:} Consider the mappings 
\[
\tilde\Phi^k:C^k(\Omega^{n-1}(M);C^\infty(M,\R))\longrightarrow
C^k(\mbox{$\bigwedge^{n-1}$}(C^\infty(M,\R));C^\infty(M,\R))
\]
defined by 
\begin{equation}\label{Phik}
\tilde\Phi^k(c^k)(F_1,\dots ,F_k)=c^k(\Phi(F_1),\dots ,\Phi(F_k)),
\end{equation}
for $F_1,\dots ,F_k\in \bigwedge^{n-1}(C^\infty(M,\R)).$

From (\ref{cF}), (\ref{c1}) and (\ref{phi}), we get
\[
\Phi(\{F_i,F_j\}')=\lcf \Phi(F_i),\Phi(F_j) \rcf.
\]
Using this fact, (\ref{ch}), (\ref{cohole}),  (\ref{rF}), (\ref{oca})
and (\ref{Phik}), we obtain that the mappings $\tilde\Phi^k$ induce a
homomorphism of complexes. 
\hfill$\Box$ 

Next, we will introduce the modular class of an oriented
Nambu-Poisson manifold. For this purpose, we prove the following result:

\begin{theorem}\label{t4.8} 
Let $(M,\Lambda)$ be an oriented $m$-dimensional
Nambu-Poisson manifold of order $n,$ with $n\geq 3,$ and $\nu$ be a
volume form. Consider the mapping 
\[
{\cal M}_{\nu}:C^\infty(M,\R) \times\dots^{(n-1}\dots\times
C^\infty(M,\R) \longrightarrow C^\infty(M,\R),
\]
defined by 
\begin{equation}\label{cm}
{\cal L}_{X_{f_1\dots f_{n-1}}}\nu={\cal M}_{\nu}(f_1,\dots
,f_{n-1}){\nu}
\end{equation}
for $f_1,\dots ,f_{n-1}\in C^\infty(M,\R)$. Then: 

\begin{enumerate}
\item
${\cal M}_\nu$ is a
skew-symmetric $(n-1)$-linear mapping and a derivation in each
argument with respect to the usual product of functions. 
Thus, ${\cal M}_{\nu}$ defines an $(n-1)$-vector on $M$.

\item
The mapping 
\begin{equation}\label{cm1}
{\cal M}_{\Lambda}:\Omega^{n-1}(M)\longrightarrow
C^\infty(M,\R)\makebox[1cm]{} \alpha\mapsto i(\alpha){\cal M}_\nu
\end{equation}
defines a $1$-cocycle in the Leibniz cohomology complex associated to
the Leibniz algebroid
$(\mbox{$\bigwedge^{n-1}$}(T^*M),$ $\lcf \;,\;\rcf, \#)$.

\item
The cohomology class $[{\cal M}_{\Lambda}]\in
H^1(\Omega^{n-1}(M);C^\infty(M,\R))$ does not depend on the chosen
volume form.
\end{enumerate}

\end{theorem}
{\bf Proof:}
$(i)$ It follows using (\ref{skew}), (\ref{rL}), (\ref{ch}),
(\ref{2.5'})  and (\ref{cm}).

\medskip

$(ii)$ We will show that for all $\alpha\in \Omega^{n-1}(M)$ we have:
\begin{equation}\label{lsv}
{\cal L}_{\#\alpha}\nu=[i(\alpha){\cal M}_\nu+
(-1)^{n-1}(i(d\alpha)\Lambda)] \nu. 
\end{equation}

Indeed, suppose that $\alpha=fdf_1\wedge \dots \wedge df_{n-1},$
with $f, f_1,\dots ,f_{n-1}\in C^\infty(M,\R).$ 

A direct computation proves that
\begin{equation}\label{Z}
\begin{array}{lcl}
{\cal L}_{\#\alpha}{\nu}&=&df\wedge i_{X_{f_1\dots f_{n-1}}}\nu+f{\cal
M}_\nu(f_1,\dots ,f_{n-1})\nu\\
&=&df\wedge i_{X_{f_1\dots f_{n-1}}}\nu + (i(\alpha){\cal M}_\nu) \nu.
\end{array}
\end{equation}

Now, since $i_{X_{f_1\dots f_{n-1}}}(df\wedge \nu)=0,$ we deduce that
\[
df\wedge i_{X_{f_1\dots f_{n-1}}}\nu=X_{f_1\dots f_{n-1}}(f)\nu.
\]
Adding this formula to (\ref{Z}) we obtain that (\ref{lsv}) holds for
$\alpha=fdf_1\wedge \dots \wedge df_{n-1}.$ But this implies that
(\ref{lsv}) holds for all $\alpha\in \Omega^{n-1}(M).$  

Using (\ref{3.8'}), (\ref{lsv}) and Theorem \ref{t3.5}, we have that  
\[
i(\lcf \alpha,\beta\rcf){\cal M}_\nu={\cal L}_{\#\lcf
\alpha,\beta\rcf}\nu+ (-1)^n(i(d\lcf \alpha,\beta \rcf)\Lambda)\nu=
\#\alpha(i(\beta){\cal
M}_{\nu})-\#\beta(i(\alpha){\cal M}_\nu).
\]
This proves $(ii)$  
(see (\ref{oca})).

\medskip

$(iii)$ Let $\nu'$ be another volume form on $M$. Then there exists
$f\in C^\infty(M,\R)$, $f\not=0$ at every point, such that $\nu'=f\nu$.
We can suppose, without the loss of generality, that $f>0.$

A direct computation, using (\ref{lsv}), shows that for all $\alpha\in
\Omega^{n-1}(M)$  
\[
i(\alpha){\cal M}_{\nu'}=i(\alpha){\cal M}_\nu + \#\alpha(lnf)
\]
which implies that (see (\ref{oca}))
\[
{\cal M}_{\nu'}={\cal M}_\nu + \partial(lnf).
\]
\hfill$\Box$ 

Theorem \ref{t4.8} allows us to introduce the following definition.

\begin{definition}\label{4.8}
Let $(M,\Lambda)$ be an oriented Nambu-Poisson
manifold of order $n,$ with $n\geq 3$, and ${\cal M}_{\Lambda}$ be
the cocycle defined by 
(\ref{cm1}). The cohomology class 
\[
[{\cal M}_\Lambda]\in
H^1(\Omega^{n-1}(M);C^\infty(M,\R))
\]
is called the modular class of 
$(M,\Lambda).$ 
\end{definition}

\begin{remark}{\rm 
Definition \ref{4.8} extends for  Nambu-Poisson
manifolds of order greater than $2$ the notion of modular class of a Poisson manifold
introduced by Weinstein \cite{We} (see also \cite{BZ}).}
\end{remark}

For a Nambu-Poisson structure induced by a volume form, we deduce:

\begin{proposition}\label{4.9}
Let $M$ be an oriented $m$-dimensional manifold and $\nu$ a volume
form on $M$.  Then the modular class of $(M,\Lambda_\nu)$ is null.
\end{proposition}

{\bf Proof:} 
Using (\ref{2.5'}), (\ref{if}) and  (\ref{vol}), we obtain that 
\[
{\cal L}_{X_{f_1\dots f_{n-1}}}\nu=0,
\]
for all $f_1,\dots ,f_{n-1}\in C^\infty(M,\R).$ This implies that
${\cal M}_\nu=0$ and therefore, ${\cal M}_{\Lambda_{\nu}}=0$. 
\hfill$\Box$

\begin{remark}\label{corop}{\rm 
Suppose that $N$ and $L$ are oriented
manifolds and let $\nu$ be a volume
form on $N$. The Nambu-Poisson structure $\Lambda_\nu$ on $N$ induces
a Nambu-Poisson structure $\Lambda$ on the product manifold
$M=N\times L$ (see Examples \ref{2.1}) and from Proposition
\ref{4.9}, it follows that 
the modular
class of $(M,\Lambda)$ is null.}
\end{remark}

Using Theorem \ref{thl} and Remark \ref{corop}, we have the following.

\begin{corollary}
Let $M$ be an oriented $m$-dimensional Nambu-Poisson manifold of
order $n$, with $3\leq n\leq m$. If at a point $x\in M$ we have
$\Lambda(x)\not=0,$ then there exists an open neighborhood $U$ of $x$
in $M$ such that the modular class of $(U,\Lambda_U)$ is null. Here
$\Lambda_U$ denotes the Nambu-Poisson structure induced by $\Lambda$
on $U.$ 
\end{corollary}

The above results and the following example show that the vanishing
of the modular class of a Nambu-Poisson manifold is closely
related with its regularity.
\begin{example}{\rm 
Consider on $\R^3$ the $3$-vector defined by 
\[
\Lambda=x^3\frac{\partial}{\partial x^1}\wedge
\frac{\partial}{\partial x^2}\wedge \frac{\partial}{\partial x^3},
\]
where $(x^1,x^2,x^3)$ denote the usual coordinates on $\R^3$.

The $3$-vector $\Lambda$ defines a
Nambu-Poisson structure of order $3$ on $\R^3$.

Let $\nu$ be the volume form given by 
\[
\nu=dx^1\wedge dx^2\wedge dx^3.
\]
A direct computation proves that 
\[
X_{x^1x^2}=x^3\frac{\partial}{\partial x^3},\makebox[1cm]{} 
X_{x^1x^3}=-x^3\frac{\partial}{\partial x^2},\makebox[1cm]{} 
X_{x^2x^3}=x^3\frac{\partial}{\partial x^1},
\]
and 
\[
{\cal L}_{X_{x^1x^2}}\nu=\nu,\makebox[1cm]{}
{\cal L}_{X_{x^1x^3}}\nu={\cal L}_{X_{x^2x^3}}\nu=0.
\]
Thus, ${\cal M}_\nu=\displaystyle\frac{\partial}{\partial x^1}\wedge
\displaystyle\frac{\partial}{\partial x^2}.$

Now, if the class modular of $(\R^3,\Lambda)$ would be null then there
exists $f\in C^\infty(\R^3,\R)$ such that 
\[
i(\alpha){\cal M}_\nu=\partial f(\alpha),
\]
for all $\alpha\in \Omega^{2}(\R^3)$. Taking $\alpha=dx^1\wedge dx^2$,
we would deduce that 
\[
1=X_{x^1x^2}(f)=x^3\frac{\partial f}{\partial x^3}.
\]
But this is not possible. Thus $[{\cal M}_\Lambda]\not=0$.}
\end{example}

\section*{Acknowledgments}

This work has been partially supported through grants DGICYT
(Spain) (Projects PB97-1257 and  PB97-1487) and Project U.P.V.
127.319-EA043/97.
ML and JCM wish to express his gratitude for the heartly
hospitality offered to them in the Department of  Fundamental Mathematics
(University of La Laguna) and in the 
Department of Mathematics (University of Basque Country) respectively,
where part of this work was done.

\end{document}